\documentclass[aps,prl,amsmath,amssymb,amsfonts,showpacs,superscriptaddress,floatfix,twocolumn]{revtex4}

\usepackage{amssymb}
\usepackage{graphics}
\usepackage{graphicx}

\newcommand{\be}{\begin{eqnarray}}
\newcommand{\ee}{\end{eqnarray}}
\newcommand{\bea}{\begin{eqnarray}}
\newcommand{\eea}{\end{eqnarray}}
\newcommand{\bma}{\begin{subequations}}
\newcommand{\ema}{\end{subequations}}

\DeclareMathOperator{\poly}{poly}
\DeclareMathOperator{\Tr}{tr}

\newcommand{\bra}[1]{\langle #1 |}
\newcommand{\ket}[1]{| #1 \rangle}

\newcommand{\set}[1]{\lbrace #1 \rbrace}
\newcommand{\RR}{\mathbb{R}}
\newcommand{\SSS}{\mathcal{S}}
\def\N{N}   % the N in N-representability problem
\def\m{d}   % number of modes (number of single particle wavefunctions, local dimension)
\def\n{N}   % the number of particles
\begin{document}

\title{$\N$-representability is QMA-complete}
\author{Yi-Kai \surname{Liu}}
\affiliation{Computer Science and Engineering, University of California, San Diego, US}
\author{Matthias \surname{Christandl}}
\affiliation{Centre for Quantum Computation, Centre for Mathematical Sciences, DAMTP, University of Cambridge, Wilberforce Road, Cambridge CB3 0WA, United Kingdom}
\author{F. \surname{Verstraete}}
\affiliation{Institute for Quantum Information, Caltech, Pasadena, US} \affiliation{Facult\"at f\"ur Physik,
Universit\"at Wien, Boltzmanngasse 5, A-1090 Wien, Austria}

\date{\today}

\begin{abstract}
%We show that the $\N$-representability problem, a central object of study in the field of quantum chemistry, is
%QMA-complete and hence NP-hard.
We study the computational complexity of the $\N$-representability problem in quantum chemistry.  We show that
this problem is QMA-complete, which is the quantum generalization of NP-complete.  Our proof uses a simple mapping
from spin systems to fermionic systems, as well as a convex optimization technique that reduces the problem of
finding ground states to $\N$-representability.
%In addition, we show that pure-state $\N$-representability is in QMA,
%and thus is no harder than the standard $\N$-representability problem.
\end{abstract}

\maketitle

The central theoretical problem in the field of many-body strongly
correlated quantum systems is to find efficient ways of simulating
Schr\"odinger's equations: it is very easy to write down those
equations, but notoriously difficult to solve them or even to
find approximate solutions. The main difficulty is the fact that the
dimension of the Hilbert space describing a system of $\N$ quantum
particles scales exponentially in $\N$. This makes a direct
numerical simulation intractable: every time an extra particle is
added to the system, the computational resources would have to be
doubled.

The situation is not hopeless, however, as in principle it could be that all
physical wavefunctions, i.e., the ones that are realized in nature, have very special properties and can
be parameterized in an efficient way. The idea would then be to propose a variational class of wavefunctions that
capture the physics of the systems of interest, and then do an optimization over this restricted class. This
approach has proven to be very successful, as witnessed by mean field theory and renormalization group
methods. So far, an efficient variational class to describe complex wavefunctions such as those arising in quantum
chemistry has not been found.

One of the basic problems in quantum chemistry is to find the ground
state of a Hamiltonian describing the many-body system of an atom or
molecule. The essential element that makes typical Hamiltonians very
ungeneric is the fact that at most 2-body interactions occur. This
implies that the number of free parameters in such Hamiltonians
scales at most quadratically in the number of particles or modes,
and hence the ground states of all such systems form a
small-dimensional manifold.

In the case of a Hamiltonian with only
2-body interactions, the energy corresponding to a wavefunction is
completely determined by its 2-body correlation functions, and as a
consequence the ground state will be the one with extremal 2-body
reduced density operators. This fact was realized a long time ago,
and led Coulson \cite{Coulson} to propose the following problem:
given a set of $\N$ quantum particles, can we characterize the
allowed sets of 2-body correlations or density operators between all
pairs of particles?

If the particles under consideration are fermions, this has been
called the \textit{$\N$-representability problem}~\cite{Coleman}.
%In the case of fermions, the complete $\N$-body wavefunction can
%always be written as a linear combination of Slater determinants,
%and one is interested in the reduced density operators acting on
%pairs of fermions.
Here, we consider the reduced density operators acting on pairs of
fermions, and we want to decide whether they are consistent with
some global state over $\N$ fermions.  %A variant of this problem is
%\textit{pure-state} $\N$-representability, where we impose an
%additional requirement that the $\N$ fermions must be in a pure
%state; this problem seems more complicated, because the set of
%feasible reduced density operators is no longer convex.

An efficient solution to the $\N$-representability problem would be a huge breakthrough, as it would (for example)
allow us to calculate the binding energies of all molecules.  Therefore, a very large  effort has been devoted to
solving this problem and there exists a large literature on this subject (see e.g. \cite{book1,book,Mazziotti}).

Here we will show that the $\N$-representability problem is intractable,
as it is QMA-complete and hence NP-hard. The complexity class QMA
(Quantum Merlin-Arthur) is the natural generalization of the class NP
(nondeterministic polynomial time) to the setting of quantum
computing. %it was introduced by Kitaev \cite{Kitaev}. %We will also
%show that pure-state $\N$-representability is in QMA, and thus is no
%harder than the standard version of the problem.
Colloquially, a problem is in QMA if there exists an efficient quantum algorithm that, when given a possible
solution to the problem, can verify whether it is correct; here the ``solution'' may be a quantum state on
polynomially many qubits~\footnote{Technically, the problem must be formulated as a yes-or-no question, e.g.,
``for a given Hamiltonian $H$, is the ground state energy less than $\gamma$?'' If the answer is yes, there exists
a ``witness'' (e.g., the ground state) which can prove it.  Moreover, there is an efficient algorithm that can
check the validity of a witness.}. A problem is QMA-hard if it is at least as hard as any other problem in QMA;
that is, given an efficient algorithm for this problem, one could solve every other problem in QMA
efficiently~\footnote{This includes search and optimization problems, e.g., ``find the ground state energy of
$H$.'' These problems may be QMA-hard, even though they are not phrased as yes-or-no questions.}. We say that a
problem is QMA-complete if it is in QMA and it is also QMA-hard.

In a seminal work, Kitaev \cite{Kitaev}
%quantized the proof of Levin and Cook where they showed that 3-SAT is NP-complete and
proved that the Local Hamiltonian problem---determining the ground state energy of a spin Hamiltonian that is a
sum of 5-body terms (on $n$ qubits), with accuracy $\pm \varepsilon$ where $\varepsilon$ is inverse polynomial in
$n$---is QMA-complete. In fact, it was later shown that this problem remains QMA-complete when restricted to
2-body interactions \cite{Kempe1,Kempe2}, and even in the case of geometrically local interactions \cite{Terhal}.

Because the Hamiltonians under consideration are local, the
corresponding ground states have extremal local properties. The dual
problem of determining the ground state energy of a local Hamiltonian
is to decide whether a given set of local
density operators can be realized as the reduced
density operators of the same global state. Checking the
consistency of such a set of local density operators $\rho^{(ij)}$,
i.e., checking whether there exists a global state $\sigma$
compatible with those local density operators, is a QMA-complete
problem itself \cite{Yi-Kai}. In the present paper, we will use very
similar techniques to prove that $\N$-representability, which is the
fermionic version of that problem, is also QMA-complete.

To sketch the main ideas of the proof, we will first consider the classical marginal problem:
%given a probability distribution
%over $\N$ discrete variables that can take the values $\pm 1$, determine the extreme points of all the 2-point
%marginal distributions
%\[p(a_i,a_j)=\sum_{k_1,...k_{i-1},k_{i+1},...k_{j-1},k_{j+1},...}p\left(a_{k_1},a_{k_2},...\right).\]
suppose we have $\N$ random variables that can take the values $\pm
1$, according to some joint probability distribution, and we define
the 2-variable marginal distributions
\[ p(a_i,a_j) = \sum_{a_1,...,a_{i-1},a_{i+1},...,a_{j-1},a_{j+1},...,a_\N} p(a_1,a_2,...,a_\N). \]
Given a set of marginal probability distributions, we would like to check whether they are consistent,
i.e., whether there exists a global probability distribution whose marginals are equal to the ones we
were given.

Suppose that there exists an efficient algorithm (whose running time is polynomial in $\N$) to solve this problem.
Then it would be possible to identify ground states of all Ising spin glasses. The strategy would be as follows:
since the set of consistent marginals is convex and the energy is a linear function of the marginals, the problem
amounts to minimizing a linear function subject to a set of convex constraints that can be checked in polynomial
time. This can be done in polynomial time using the ellipsoid method \cite{GLS}. It therefore follows that an
efficient algorithm for checking consistency allows to find ground states of Ising spin glasses, a problem that is
known to be NP-hard \cite{Barahona}. Hence the problem of determining whether a set of binary marginals is
consistent is itself NP-hard.
% Note: finding ground states of Ising models is not in NP because it's a search problem, not a decision problem.
% Also, consistency of marginals is not in NP because it's not clear how to represent the global probability
% distribution as a classical witness; however, the problem is in MA.

Let us now formulate the $\N$-representability problem in the
context of quantum chemistry. Electrons and nuclei tend to arrange
themselves such as to minimize their energy, and the binding energy
of a molecule can be determined by calculating the minimal energy of
the corresponding Hamiltonian. In practice, the nuclei are fairly
well localized and can be treated as classical degrees of freedom,
and the wavefunction of the $\n$ electrons can be approximated as a
linear combination of tensor products of the $\m$
single-particle modes in the system (those form the basis set).

As electrons are fermionic, the complete wavefunction must be
antisymmetric, and this is most easily taken into account by working
in the formalism of second quantization:
\[
|\psi\rangle = \sum_{{\tiny\begin{array}{l}i_1,...,i_\m=0\\
i_1+...+i_\m=\n \end{array}}}^1
c_{i_1,...,i_\m} (a_1^\dagger)^{i_1}...(a_\m^\dagger)^{i_\m} |\Omega\rangle.
\]
Here $a_j^\dagger$  is the creation operator for the $j$'th mode,
and $|\Omega\rangle$ represents the vacuum state without fermions.
The creation and annihilation operators obey the following
anticommutation relations:
\[
\{a_i,a_j\} = 0 = \{a_i^\dagger,a_j^\dagger\} \hspace{1cm}
\{a_i,a_j^\dagger\} = 2\delta_{ij}.
\]
Note that we restrict ourselves to the subspace of states with exactly
$\n$ fermions.  $\m$ denotes the number of modes, which is typically
much larger than $\n$. The number of degrees of freedom is $\binom{\m}{\n}$,
which grows exponentially in $\n$ when $\m \geq c\n$ for some constant $c>1$.

In the case of quantum chemistry, the Hamiltonian typically contains
only one- and two-body interactions between all modes, and so it can
be written as a linear combination of terms of the form $a_i^\dagger
a_j$ and $a_i^\dagger a_j^\dagger a_l a_k$.
%and the energy is therefore a linear combination of terms
%\bea \rho^{(1)}_{ij}&\equiv&\langle\psi|a_i^\dagger a_j|\psi\rangle\\
%\rho^{(2)}_{ijkl}&\equiv&\langle\psi|a_i^\dagger a_j^\dagger a_ka_l|\psi\rangle\label{cons}\eea

The 2-fermion reduced density matrix (2-RDM) is calculated by
tracing out all but two of the fermions:
\[
\rho^{(2)} = \Tr_{3,\ldots,\n} \rho^{(N)}.
\]
where $\rho^{(N)}$ is a mixture of states $\ket{\psi}$ with exactly $\n$ fermions. In the language of second
quantization, the matrix elements of the 2-RDM are given by:
\[
\rho^{(2)}_{ijkl} = \frac{1}{\n(\n-1)} \langle a_k^\dagger a_l^\dagger a_j a_i \rangle.
\]

The $\N$-representability problem (with $\m$ modes) can now be
stated as follows. Consider a system of $\n$ fermions and $\m$ modes,
$\m \leq \poly(\n)$. (For purposes of complexity, we consider $\n$
to be the ``size'' of the problem.) We are given a 2-fermion density
matrix $\rho$, of size $\frac{\m(\m-1)}{2} \times
\frac{\m(\m-1)}{2}$, where each entry is specified with
$\poly(\n)$ bits of precision.  In addition, we are given a real
number $\beta \geq 1/\poly(\n)$, specified with $\poly(\n)$ bits of
precision. The problem is to distinguish between the following two
cases:
\begin{itemize}
\item There exists an $\n$-fermion state $\sigma$ such that
$\Tr_{3,\ldots,\n}(\sigma) = \rho$.  In this case, answer ``YES.''
\item For all $\n$-fermion states $\sigma$,
$\lVert \Tr_{3,\ldots,\n}(\sigma) - \rho \rVert_1 \geq \beta$.
In this case, answer ``NO.''
\end{itemize}
If neither of these cases applies, then one may answer either ``YES'' or ``NO.''
Note that we do not insist on solving the problem exactly; we allow
an error tolerance of $\beta \geq 1/\poly(\n)$. We use the $\ell_1$
matrix norm or trace distance, $\lVert A \rVert_1 = \Tr |A|$, to
measure the distance between $\sigma$ and $\rho$.

%We define the pure-state $\N$-representability problem in a similar
%way, with the following modifications.  In addition to $\rho$ and
%$\beta$, we are given a real number $\delta \geq 1/\poly(\n)$,
%specified with $\poly(\n)$ bits of precision.  We have to distinguish
%between these two cases:
%\begin{itemize}
%\item There exists an $\n$-fermion pure state $\sigma$ such that
%$\Tr_{3,\ldots,\n}(\sigma) = \rho$.  In this case, answer ``YES.''
%\item For all $\n$-fermion states $\sigma$ such that $\Tr(\sigma^2)
%\geq 1-\delta$, we have that
%$\lVert \Tr_{3,\ldots,\n}(\sigma) - \rho \rVert_1 \geq \beta$.
%In this case, answer ``NO.''
%\end{itemize}
%Note that we use $\Tr(\sigma^2)$ to measure the purity of the state
%$\sigma$, and we allow an error tolerance $\delta \geq 1/\poly(\n)$.

We will show that $\N$-representability is QMA-complete. The proof consists
of two parts.  First, we show that any 2-local Hamiltonian of spins
can be simulated using a 2-local Hamiltonian of fermions with
$\m=2\n$. Using techniques of convex programming, we show that an
oracle for $\N$-representability would allow us to estimate the
ground state energies of 2-local Hamiltonians; thus,
$\N$-representability is QMA-hard.

Second, we show that
$\N$-representability is in QMA; specifically, we construct a
quantum verifier that can check whether a 2-particle state is
$\N$-representable, given a suitable witness.
%by using the fact that QMA+=QMA \cite{Aharonov} and that consistency of
%local density operators of a fermionic state of $\m$ modes can be checked by calculating many-body expectation values
%on a $N$-qubit state obtained by performing a Jordan-Wigner transformation.
%We also show that pure-state $\N$-representability is in QMA, using similar techniques.

\vskip 10pt

Let us first show how to map a 2-local Hamiltonian, defined on a system of $\n$ qubits, to a 2-local Hamiltonian
on fermions, with $\m = 2\n$ modes; this is the opposite of what has been done in \cite{VC}. The idea is to
represent each qubit $i$ as a single fermion that can be in two different modes $a_i,b_i$; so each $\n$-qubit
basis state corresponds to the following $\n$-fermion state:
\begin{equation}
\ket{z_1} \otimes \cdots \otimes \ket{z_\n} \mapsto
   (a_1^\dagger)^{1-z_1} (b_1^\dagger)^{z_1} \cdots
   (a_\n^\dagger)^{1-z_\n} (b_\n^\dagger)^{z_\n} \ket{\Omega}.
\label{map0}
\end{equation}
Also, all the relevant single-qubit operators should correspond to
bilinear functions of the creation and annihilation operators (this
construction guarantees that operators on different qubits commute).
This can be achieved by the following mapping:
\begin{equation}
\sigma^x_i \leftrightarrow a_i^\dagger b_i + b_i^\dagger a_i, \
\sigma^y_i \leftrightarrow i\left( b_i^\dagger a_i - a_i^\dagger b_i \right), \
\sigma^z_i \leftrightarrow \openone - 2 b_i^\dagger b_i.
\label{map}
\end{equation}
%\begin{eqnarray}
%\sigma^x_i &\leftrightarrow& a_i^\dagger b_i + b_i^\dagger a_i \nonumber\\
%\sigma^y_i &\leftrightarrow& i\left( b_i^\dagger a_i - a_i^\dagger b_i \right) \nonumber\\
%\sigma^z_i &\leftrightarrow& \openone - 2 b_i^\dagger b_i \label{map}
%\end{eqnarray}
It can easily be checked that these fermionic operators obey the
Pauli commutation relations within the subspace spanned by states
of the form (\ref{map0}).

Recall that the qubit Hamiltonian can be written as a linear combination
of Pauli operators.  So the fermionic Hamiltonian can now be created
by rewriting these Pauli operators with respect
to the above mapping (note that only bilinear terms occur, and hence
operators on different sites commute). Notice that if the qubit
Hamiltonian only contains 2-body terms, then the fermionic
Hamiltonian will only contain terms with at most 2 annihilation and
2 creation operators.

The only thing that is left to do is to guarantee that there is exactly one fermion on every site $i$.
This can be achieved by adding the following projectors as extra terms in the fermionic Hamiltonian:
\[P_i=(2a_i^\dagger a_i-\openone)(2b_i^\dagger b_i-\openone).\]
All the $P_i$ are biquadratic and commute with all the operators introduced
in (\ref{map}), and hence the complete Hamiltonian will be block
diagonal. By making the weights of these projectors large enough (a
constant times the norm of the Hamiltonian, which is at
most polynomial in $\n$), we can always guarantee that the ground
state of the full Hamiltonian will have exactly one fermion per site
\footnote{Another option would be to represent each qubit $i$
using either zero or two fermions, e.g., for a single qubit,
$\ket{0}$ and $\ket{1}$ correspond to $\ket{\Omega}$ and
$a_i^\dagger b_i^\dagger \ket{\Omega}$. Define the mapping
$\sigma^x_i \leftrightarrow  \left( a_i^\dagger b_i + b_i^\dagger a_i \right) +
                             \left( b_i a_i + a_i^\dagger b_i^\dagger \right)$,
$\sigma^y_i \leftrightarrow i\left( b_i^\dagger a_i - a_i^\dagger b_i \right) +
                            i\left( a_i^\dagger b_i^\dagger - b_i a_i \right)$,
$\sigma^z_i \leftrightarrow \openone - 2 b_i^\dagger b_i$.
Then the Hamiltonian would act identically on the subspace with one fermion per site,
and on the subspace with zero or two fermions per site,
%Then the subspace with even fermion occupation number would be isomorphic to the one with odd occupation number,
and even a small constant in front of the projectors $P_i$ would be enough to guarantee
that we end up in the right subspace.}.

\vskip 10pt

Let us now assume that we have an efficient algorithm for
$\N$-representability.  We claim that this allows us to efficiently
determine the ground state energy of a 2-local Hamiltonian on
qubits, a problem which is known to be QMA-hard \cite{Kempe2}.  We
start by transforming the Hamiltonian $H_{\text{qubit}}$ on $\n$
qubits with 2-particle interactions into a Hamiltonian
$H_{\text{fermi}}$ on $\m=2\n$ fermionic modes with 2-particle
interactions (as described above).  The problem of finding the
ground state energy of $H_{\text{fermi}}$ can be expressed as a
convex program, which can be solved in polynomial time using an
oracle for $\N$-representability (this is what we will show below).
Our approach is similar to \cite{Yi-Kai}, though in this case we
are dealing with fermions rather than qubits.

The basic idea is to construct a convex program that finds a
2-particle density matrix $\rho$ which is $\N$-representable, and
which minimizes the expectation value of $H_{\text{fermi}}$.  This
program has polynomially many variables, and it is easy to see that
the set of $\N$-representable states is convex, and $\langle
H_{\text{fermi}} \rangle$ is a linear function of $\rho$.

There are two minor complications.  First, $H_{\text{fermi}}$
describes a system with $2\n$ modes and an arbitrary number of
particles, whereas $\N$-representability pertains to a system with
exactly $\n$ particles.  But this is not a problem, because the
interesting behavior in $H_{\text{fermi}}$ occurs in a subspace
where all the states have exactly $\n$ particles.  So, in our convex
program, we only need to consider states $\rho$ that have exactly
$\n$ particles.

This lets us simplify $H_{\text{fermi}}$ as follows.  Since we are
only interested in how it acts on $\n$-particle states,
we can write $a_i^\dagger a_j = \frac{1}{\n-1} a_i^\dagger
(\sum_k a_k^\dagger a_k) a_j$.  So we can assume that all the terms
in $H_{\text{fermi}}$ are of the form $a_i^\dagger a_j^\dagger a_l a_k$.

Second, convex optimization algorithms usually require that the set
$K$ of feasible solutions be full-dimensional, i.e., $K$ cannot lie
in a lower-dimensional subspace.  So we have to represent the state
$\rho$ in such a way that there are no redundant variables.

Let $\SSS$ be a complete set of 2-particle observables
\footnote{One possible set of observables is the following:
First, define $a_I = a_{i_2} a_{i_1}$, for all pairs
$I = \set{i_1,i_2}$, $i_1<i_2$. Also fix an
ordering on the pairs $I$.  We now define the following observables:
$X_{IJ} = a_I^\dagger a_J + a_J^\dagger a_I$, for all $I \prec J$;
$Y_{IJ} = -i a_I^\dagger a_J +i a_J^\dagger a_I$, for all $I \prec
J$; and $Z_I = a_I^\dagger a_I$, for all $I$ except the last one.
These operators are Hermitian, with eigenvalues in the interval
$[-1,1]$. Taking real linear combinations, these operators form a
basis for the space of 2-fermion density matrices.},
and let $\ell = |\SSS|$; note that $\ell \leq \poly(\m)
\leq \poly(\n)$.  We represent $\rho$
in terms of its expectation values $\alpha_S = \Tr(S\rho)$ for all
$S \in \SSS$; let $\vec{\alpha} \in \RR^\ell$ denote the vector of
these expectation values.  We define $K$ to be the set of all
$\vec{\alpha}$ such that the corresponding state $\rho$ is
$\N$-representable.  Note that the $\N$-representability oracle
allows us to test whether a given point $\vec{\alpha}$ is in $K$.

We write our Hamiltonian in the form $H = \sum_{S \in \SSS} \gamma_S
S$ (plus a constant term); the coefficients $\gamma_S$ can be
computed using the Gram-Schmidt procedure.  It is easy to see that
$\Tr(H\rho) = \sum_{S \in \SSS} \gamma_S \alpha_S$.  Our convex
program is as follows:  find some $\vec{\alpha} \in K$ that
minimizes $f(\vec{\alpha}) = \sum_{S \in \SSS} \gamma_S \alpha_S$.

We can solve this convex program in polynomial time, using the
shallow-cut ellipsoid algorithm (see \cite{GLS}, and references
cited therein).  We mention a few technical details.  The algorithm
requires some additional information about $K$, namely a guarantee
that $K$ is contained in a ball of radius $R$ centered at 0, and $K$
contains a ball of radius $r$ centered at some point $p$.  (Also, the
running time of the algorithm grows polynomially in $\log(R/r)$.) In
our case, we can set $R = \sqrt{\ell}$, and $r = 1/\poly(\ell)$
\footnote{It is easy to see that $K$ is contained in a ball
of radius $R = \sqrt{\ell}$, since for all $\vec{\alpha} \in K$,
we have $-1 \leq \alpha_S \leq 1$.
\vskip 1pt \hspace{4pt}
The second claim, that $K$ contains a
ball of radius $r = 1/\poly(\ell)$, is less trivial.  The proof is as
follows.
\vskip 1pt \hspace{4pt}
We will consider $\N$-representability for different values of $\N$;
let $K_{\n}$ denote the set of all vectors $\vec{\alpha}$ that are
$\n$-representable.  Obviously, $K_2$ contains a ball of radius
$1/\poly(\ell)$ (this is the trivial case).  We will prove that
$K_{\n}$ contains a ball of radius $1/\poly(\ell)$, for all
$2 \leq \n \leq \m-2$.
\vskip 1pt \hspace{4pt}
We define ``particle-hole'' observables, by replacing $a_i$ with
$a_i^\dagger$, and vice versa:
$X'_{IJ} = a_I a_J^\dagger + a_J a_I^\dagger$; $Y'_{IJ} = -i a_I
a_J^\dagger +i a_J a_I^\dagger$; and $Z'_I = a_I a_I^\dagger$.  Let
$\vec{\alpha}'$ denote a vector containing expectation values for these
observables; let $K'_{\n}$ be the set of all $\vec{\alpha}'$ that are
$\n$-representable.
\vskip 1pt \hspace{4pt}
First, we claim that $K_2 = K'_{\m-2}$.  Notice that there is a
natural correspondence between the 2-particle Slater basis states
and the $(\m-2)$-particle Slater basis states:  the 2-particle state
with modes $i$ and $j$ occupied corresponds to the $(\m-2)$-particle
state with modes $i$ and $j$ empty.
\vskip 1pt \hspace{4pt}
So take any point $\alpha \in K_2$, which represents the expectation
values of the 2-particle observables for some 2-particle state $\sigma$.
Use $\sigma$ to construct a $(\m-2)$-particle state $\tau$, by
replacing each 2-particle Slater basis state with the corresponding
$(\m-2)$-particle Slater basis state.  Then the expectation values
of the 2-particle observables for $\sigma$ are exactly the expectation
values of the 2-hole observables for $\tau$.  So $\alpha$ is in
$K'_{\m-2}$.  This shows that $K_2 \subseteq K'_{\m-2}$.  A similar
argument shows that $K'_{\m-2} \subseteq K_2$.  So $K_2 = K'_{\m-2}$.
\vskip 1pt \hspace{4pt}
Next, we claim that there is an invertible linear transformation $A$
that maps $K'_{\m-2}$ to $K_{\m-2}$.
\vskip 1pt \hspace{4pt}
First we map $K'_{\m-2}$ to $K_{\m-2}$.  Observe that we can write
each 2-particle operator as a linear combination of 2-hole operators.
(This holds provided we restrict the operators to act only on the
subspace of $(\m-2)$-particle states.)  For instance, when $I \cap J
= \emptyset$, $a_I^\dagger a_J = a_J a_I^\dagger$.  When
$|I \cap J| = 1$, we write equations such as
$a_i^\dagger a_l^\dagger a_l a_j
 = a_i^\dagger a_j a_l^\dagger a_l
 = -a_j a_i^\dagger (1 - a_l a_l^\dagger)
 = -a_j a_i^\dagger + a_j a_i^\dagger a_l a_l^\dagger
 = -a_j a_i^\dagger + a_l a_j a_i^\dagger a_l^\dagger$;
then use the fact that
$a_j a_i^\dagger
 = (\sum_{k \neq i,j} a_k a_k^\dagger) a_j a_i^\dagger
 = \sum_{k \neq i,j} a_k a_j a_i^\dagger a_k^\dagger$.
When $I = J$, we have $a_I^\dagger a_I = \sum_{L \cap I = \emptyset}
a_L a_L^\dagger$.  (For each of these equations, it is easy to check
that the left and right sides act identically on all $(\m-2)$-particle
Slater basis states.)
\vskip 1pt \hspace{4pt}
Thus, for any $(\m-2)$-particle state $\sigma$, the expectation values
of the 2-particle observables are linear functions of the expectation
values of the 2-hole observables.  Thus we have a linear
transformation $A$ that maps $K'_{\m-2}$ to $K_{\m-2}$.
\vskip 1pt \hspace{4pt}
Similarly, we can map $K_{\m-2}$ to $K'_{\m-2}$.  We write each
2-hole operator as a linear combination of 2-particle operators (again,
restricting the operators to act only on the subspace of $(\m-2)$-particle
states).  For instance, when $I \cap J = \emptyset$, $a_I a_J^\dagger
= a_J^\dagger a_I$.  When $|I \cap J| = 1$, we write equations such as
$a_l a_i a_j^\dagger a_l^\dagger
 = a_i a_j^\dagger a_l a_l^\dagger
 = -a_j^\dagger a_i (1 - a_l^\dagger a_l)
 = -a_j^\dagger a_i + a_j^\dagger a_i a_l^\dagger a_l
 = -a_j^\dagger a_i + a_j^\dagger a_l^\dagger a_l a_i$;
then use the fact that
$a_j^\dagger a_i
 = (\frac{1}{\m-3} \sum_{k \neq i,j} a_k^\dagger a_k) a_j^\dagger a_i
 = \frac{1}{\m-3} \sum_{k \neq i,j} a_j^\dagger a_k^\dagger a_k a_i$.
When $I = J$, we write
$a_I a_I^\dagger
 = a_{i_1} a_{i_1}^\dagger a_{i_2} a_{i_2}^\dagger
 = (1 - a_{i_1}^\dagger a_{i_1}) (1 - a_{i_2}^\dagger a_{i_2})
 = 1 - a_{i_1}^\dagger a_{i_1} - a_{i_2}^\dagger a_{i_2} + a_I^\dagger a_I$;
then use the fact that $1 = \binom{\m-2}{2}^{-1} \sum_L a_L^\dagger a_L$,
and $a_i^\dagger a_i
 = \frac{1}{\m-3} \sum_{l \neq i} a_l^\dagger a_l a_i^\dagger a_i
 = \frac{1}{\m-3} \sum_{l \neq i} a_i^\dagger a_l^\dagger a_l a_i$.
\vskip 1pt \hspace{4pt}
Thus, for any $(\m-2)$-particle state $\sigma$, the expectation
values of the 2-hole observables are linear functions of the
expectation values of the 2-particle observables.  Thus we have a
linear transformation $B$ that maps $K_{\m-2}$ to $K'_{\m-2}$, and
$B = A^{-1}$.
\vskip 1pt \hspace{4pt}
We claim that $K_{\m-2}$ contains a ball of radius $1/\poly(\ell)$.
We know that $K'_{\m-2}$ contains a ball of radius $1/\poly(\ell)$
(since $K_2 = K'_{\m-2}$),
and we will show that the map $A$ does not shrink this too much.
Write the singular value decomposition $A = UDV$, where $U$ and $V$
are unitary, and $D$ is diagonal, with diagonal entries $D_{ii}>0$.
Let $B = A^{-1}$.  Looking at the matrix elements of $B$, we can see
that $\Tr(B^\dagger B) = \sum_{ij} |B_{ij}|^2 \leq \poly(\ell)$.
At the same time, $\Tr(B^\dagger B) = \Tr(U D^{-1} V V^{-1} D^{-1} U^{-1})
= \Tr(D^{-2}) \geq D_{ii}^{-2}$, for all $i$.  So we have $D_{ii} \geq
1/\poly(\ell)$, for all $i$.  Thus the map $A$ shrinks the
radius of the ball by at most a $\poly(\ell)$ factor.  So $K_{\m-2}$
contains a ball of radius $1/\poly(\ell)$.
\vskip 1pt \hspace{4pt}
Now we are almost done.  Notice that $K_{\n-1} \supseteq K_{\n}$,
for all $3 \leq \n \leq \m-2$ (since, if $\vec{\alpha}$ is consistent
with some $\n$-fermion state $\sigma$, then it is consistent with the
$(\n-1)$-fermion state $\sigma'$ that results from tracing out the
last particle).  So we conclude that $K_{\n}$ contains a ball of
radius $1/\poly(\ell)$, for all $2 \leq \n \leq \m-2$.  This completes
the proof.
}.

There is also the issue of numerical precision.  Our oracle for
$\N$-representability has limited accuracy---it may give incorrect
answers for points near the boundary of $K$.  On the other hand, we
only need an approximate solution to the convex program, in order to
solve the Local Hamiltonian problem. It turns out that $1/\poly(\n)$
precision is sufficient. The ellipsoid algorithm still works in this
setting; see \cite{GLS} for details.

Note that, in place of the ellipsoid algorithm, we could have used a
different algorithm based on random walks in convex bodies
\cite{BV}; this was the approach used in \cite{Yi-Kai}.
However, it is not clear if we could use one of the
interior-point methods for convex optimization; these methods are
substantially faster, but they usually require an explicit
description of the constraints, not just a membership oracle.

This completes the proof that $\N$-representability is QMA-hard.
As a corollary, we have also proven that estimating the ground state energy for
local fermionic Hamiltonians is QMA-hard. Note that the use of
the ellipsoid algorithm to reduce a convex optimization problem to a
convex membership problem is not new; see \cite{GLS} for references
to related work.

\vskip 10pt

Next, we show that $\N$-representability is in QMA.  That is, we
construct a poly-time quantum verifier $V$ that takes two inputs:
a description of the problem (that is, $\rho$ and $\beta$); and a
``witness'' $\tau$, which is a quantum state on polynomially many qubits.
The verifier $V$ should have the following property:
if $\rho$ is $\N$-representable, there exists a witness $\tau$
that causes $V$ to output ``true'' with probability $\geq p_1$;
if $\rho$ is not $\N$-representable (within error tolerance $\beta$),
then for all possible states $\tau$, $V$ outputs ``true'' with
probability $\leq p_0$; and $p_1 - p_0 \geq 1/\poly(\n)$.

The idea is that, when $\rho$ is $\N$-representable, the correct witness
$\tau$ consists of (multiple copies of) an $\n$-fermion state $\sigma$
that satisfies $\Tr_{3,\ldots,\n}(\sigma) = \rho$.  Then the verifier
can use quantum state tomography to compare $\sigma$ and $\rho$.

%Say we have a 2-fermion state $\rho$ which is $\N$-representable;
%then it is natural to use the corresponding $\n$-fermion state
%$\sigma$ as the witness.

We represent the $\n$-fermion state $\sigma$ using $\m$ qubits, via
the following mapping:
\[
(a_1^\dagger)^{i_1} ... (a_\m^\dagger)^{i_\m} \ket{\Omega}
\leftrightarrow \ket{i_1} \otimes ... \otimes \ket{i_\m}.
\]
Call the resulting qubit state $\tilde{\sigma}$.  We use the Jordan-Wigner
transform to map the fermionic annihilation operators to qubit operators:
\[
a_i \leftrightarrow A_i \equiv -(\otimes_{k<i}\hspace{.1cm}\sigma_k^z) \otimes \ket{0} \bra{1}_i.
\]
Thus, an observable $O = a_i^\dagger a_j^\dagger a_l a_k +
a_k^\dagger a_l^\dagger a_j a_i$ is transformed into $\tilde{O} =
A_i^\dagger A_j^\dagger A_l A_k + A_k^\dagger A_l^\dagger A_j A_i$,
which is a tensor product of many single-qubit observables and one
four-qubit observable.

We claim that the expectation value $\langle
\tilde{O} \rangle$ can be estimated efficiently.  Without loss of
generality, assume that the eigenvalues of $\tilde{O}$ lie in the
interval $[0,1]$.  Then there is an efficiently-implementable
measurement which outputs ``1'' with probability $\langle
\tilde{O} \rangle$ \footnote{For example, let $\lambda_i$ and
$\ket{\theta_i}$ be the eigenvalues and eigenvectors of $\tilde{O}$,
and note that they are easy to compute.  Add an ancilla qubit,
perform the unitary operation $U:\: \ket{\theta_i}\ket{0} \mapsto
\ket{\theta_i} \left( \sqrt{1-\lambda_i}\ket{0} +
\sqrt{\lambda_i}\ket{1} \right)$, and measure the ancilla in the 0,1
basis.}.  By repeating this measurement on multiple copies of the
same state, we can estimate $\langle \tilde{O} \rangle$; for our
purposes, it is enough to have polynomially many copies of the state.

We now describe the verifier $V$.  The witness $\tau$ consists of
several (i.e., polynomially many) blocks, where each block has $\m$ qubits,
supposedly representing one copy of the state $\tilde{\sigma}$.  On each
block, $V$ measures the observable $\sum_k \ket{1} \bra{1}_k$,
and if the outcome does not equal $\n$, $V$ outputs ``false.''
This projects each block onto the space of $\n$-particle states.

Next, $V$ performs measurements on each block, to estimate the expectation values of $\tilde{\sigma}$, for a
suitable set of observables.  Then $V$ checks whether they match the expectation values of $\rho$.  One problem
arises:  the prover could try to cheat by entangling the different blocks of qubits.  One can show that this does
not fool the verifier, using a Markov argument, as was done in \cite{Aharonov}.  This suffices to show that
$\N$-representability is in QMA, and hence finishes the proof.

What can be said about the complexity of the pure-state $\N$-representability-problem, where one has to decide
whether the reduced density operators arise from a pure state of N fermions? In that case, the verifier must be
able to convince himself that the state he gets is pure. This can be done when he gets two states $\rho$ and
$\sigma$ that are promised to be uncorrelated, i.e. that he gets the state $\rho\otimes\sigma$: then Arthur can
calculate the expectation value of the observable $\Tr(\rho\sigma)$, and this can only be close to $1$ when $\rho$
is pure and $\sigma\simeq\rho$. Indeed, if $\Tr(\sigma^2) \leq 1-\varepsilon$, then for all states $\tau$,
$\Tr(\sigma\tau) \leq \sqrt{\Tr(\sigma^2)\Tr(\tau^2)} \leq (1-\varepsilon)^{1/2} \leq 1-\varepsilon/2$. The
problem however is that Merlin can cheat, and hand out a correlated state to Arthur, such that the above test
becomes inconclusive. This is precisely the feature that distinguishes the complexity class QMA(k) from QMA
\cite{Matsumoto}: in QMA(k), the verifier is promised to get a tensor product of different states, and it has been
conjectured that QMA(k) is strictly larger than QMA. The above discussion shows that the pure-state
$\N$-representability problem is contained in QMA(k), and it is hence plausible that it is harder than the
$\N$-representability problem. It would be very interesting to investigate whether the problem is QMA(k)-complete.

\vskip 10pt

It is remarkable that checking consistency of 2-body reduced density
operators of fermionic states is so hard, while checking consistency
of 1-body reduced density operators is simple \cite{Coleman}. This
can be easily understood from the previous discussion: the extreme
points of the convex set of 1-body density operators $\langle
a_i^\dagger a_j\rangle$ correspond to ground states of Hamiltonians
only containing bilinear terms in $a_i^\dagger$ and $a_j$; such
Hamiltonians can easily be diagonalized as they represent systems of
free fermions, and hence the consistency problem can easily be
solved. As shown in \cite{Coleman}, consistency can be decided in
that case based solely on the eigenvalues of the
reduced density operators. A number of related problems, where
consistency only depends on the eigenvalues, have been
investigated recently~\cite{reduced}; most remarkably, a characterisation
has been obtained for the polytope of 1-particle marginals that are
$\N$-representable under the condition that the $\N$-particle
wavefunction must be pure \cite{Klyachko05}.

These results have to be contrasted with our problem of finding the $N$-representable 2-body density operators,
where the eigenvalues alone are not enough to decide consistency but also the eigenvectors are relevant. Actually,
let us consider the simpler problem of deciding $\N$-representability of 2-fermion density operators where only
the diagonal elements $D_{ij}=\langle a_i^\dagger a_j^\dagger a_j a_i\rangle$ are specified. If we consider the
case $\m=2\n$ and the mapping discussed above, one easily finds that the extreme points of this set would be
obtained by ground states of local spin Hamiltonians which only contain commuting $\sigma^z$ operators. These
correspond to spin-glasses, and so the problem of deciding $\N$-representability of $\{D_{ij}\}$ is NP-hard
\cite{Barahona}. It was indeed pointed out a long time ago that $\N$-representability restricted to the diagonal
elements is equivalent to a combinatorial problem \cite{Kuhn} that was later shown to be equivalent to the NP-hard
problem of deciding membership in the boolean quadric polytope \cite{Deza}.

Let's finally discuss the relevance of the above results in the context of quantum chemistry. We have shown that
it is a hopeless task to determine the ground states of all local fermionic Hamiltonians, and in particular, that
an approach by means of the $\N$-representability problem is intractable, even on a quantum computer. But it is
possible that other physical systems, e.g.~systems with different particle statistics, additional symmetry or a
limited number of modes, might allow for an efficient characterisation of the two-particle reduced density
matrices, and hence an efficient calculation of the ground states of local Hamiltonians. This seems to be the case
in e.g. one-dimensional translational invariant spin systems, where the density matrix renormalization group
\cite{DMRG} allows for a systematic approximation of the allowed convex set of reduced density operators from
within \cite{VC06}. A very different numerical approach has been developed in the context of quantum chemistry and
it known as the the contracted Schr{\"o}dinger equation~\cite{book}. In that method, one approximates the convex
set from the outside, and the $\N$-representability problem is a crucial ingredient of the algorithm. The
foregoing discussion shows that this approach has to break down in the most general case, and it would be very
interesting to investigate the conditions under which those approximations are justified.

In conclusion, we investigated the problem of $\N$-representability, and characterized its computational
complexity by showing that it is QMA-complete. Obviously, the theory of quantum computing was a prerequisite to
pinpoint the computational complexity of this classic problem. .

\vskip 10pt

\noindent \textit{Acknowledgements:} Y.K.L. and M.C. thank the Institute for Quantum Information for its
hospitality. Y.K.L. is supported by an ARO/DTO Quantum Computing Graduate Research Fellowship. M.C. acknowledges
an EPSRC Postdoctoral and a Nevile Research Fellowship which he holds at Magdalene College Cambridge, and is
supported by the EU under the FP6-FET Integrated Project SCALA, CT-015714. F.V. is supported by the Gordon and
Betty Moore Foundation through Caltech's Center for the Physics of Information, and by the National Science
Foundation under Grant No. PHY-0456720.

\end{document}